\documentclass[a4paper,12pt]{article}
\usepackage[utf8]{inputenc}
\usepackage[margin=1.2 in]{geometry}
\usepackage{cite}
\usepackage{braket}
\usepackage{amsmath}
\usepackage{amssymb}
\usepackage{amsfonts}
\usepackage{hyperref}
\usepackage{graphicx}
\usepackage{comment}
\usepackage{xcolor}
\usepackage{authblk}

\DeclareUnicodeCharacter{200D}{\,}
\DeclareUnicodeCharacter{202F}{\,}
\DeclareUnicodeCharacter{03BC}{\,} 

\title{\Huge The hierarchy problem and the vacuum stability
in two-scalar dark matter model}
\author{ \small Zohre Habibolahi$^*$, Karim Ghorbani$^*$ and Parsa Ghorbani$^\dagger$}
\date{}
\affil{\it $^*$Physics Department, Faculty of Sciences, Arak University, Arak 38156-8-8349, Iran}
\vspace{-0.6cm}
\affil{\it $^\dagger$Physics Department, Faculty of Science, Ferdowsi University of Mashhad, Iran}

\newcommand{\be}{\begin{equation}}
\newcommand{\ee}{\end{equation}}
\newcommand{\bs}{\begin{split}}
\newcommand{\es}{\end{split}}

\begin{document}

%\twocolumn

% \begin{titlepage}
% \begin{flushright}
% \end{flushright}
% \vfill
% \begin{center}
% {\Large\bf Running of the scalar couplings and vacuum stability in two-scalar dark matter model}
% \vfill
%  {\bf Zohre Habibolahi$^1$, Karim Ghorbani$^{1}$, Parsa Ghorbani$^{2}$}\\[1cm]
% {$^1$Physics Department, Faculty of Sciences, Arak University, Arak 38156-8-8349, Iran} \\
% % {$^2$Physics Department, Faculty of Science, Ferdowsi University of Mashhad, Iran}
% \end{center}
% \vfill

\maketitle
\begin{abstract}
We consider an extension to the Standard Model (SM) with two extra real singlet scalars which interact with the SM Higgs particle.  The lighter scalar is taken as the dark matter (DM) candidate.
We show that the model successfully explains the relic abundance of the DM in the universe and evades the strong bounds from direct detection experiments while respecting the perturbativity and the vacuum stability conditions. In addition, we study the hierarchy problem within the Veltman approach by solving the renormalization group equations at one-loop. We demonstrate that the addition of the real singlet scalars contributes to the Veltman parameters which in turn results in satisfying the Veltman conditions much lower than the Planck scale $ \Lambda_\text{Pl}$ down to the electroweak scale. Therefore, the presence of the extra scalars solves the fine-tuning problem of the Higgs mass. 
For the case of the two-scalar DM model we find two representative points in the 
viable parameter space which satisfy also the Veltman conditions at $\Lambda = 300$ GeV and 
$\Lambda = 1$ TeV.

% The parameter space of the model is constrained by imposing theoretical constraints including, positivity,
% vacuum stability and perturbativity. The viable parameter space is then found after we put limits
% from observed dark matter density and direct detection experiments.
% With the assumption that there is no more fermions to be added in the Standard Model, we study the hierarchy problem in the presence of the singlet scalars by utilizing the Veltman conditions.
% To this end, we first compute the $\beta$-functions of the couplings and then study their runnings in the viable parameter space. In the viable parameter space,
% we find that it is possible to satisfy the Veltman conditions well below the Planck scale and at the same time having the vacuum stability of potential.
\end{abstract}

\newpage

\tableofcontents
\section{Introduction}
\label{int}
The unnatural Higgs mass of $\sim 125$ GeV \cite{CMS:2020xrn,ATLAS:2018tdk}, yet consistent with the electroweak precision test \cite{Ciuchini:2013pca}, has been one of the driving forces behind developing models beyond the Standard Model (SM) of particle physics. 
In fact, the SM Higgs mass determines the scale of the electroweak 
symmetry breaking, and on the other hand, the amount of the Higgs mass
strongly depends on the quadratic divergences from the loop corrections.
In order to get a negligible one loop correction within the SM (known as the Veltman condition), Veltman found that the Higgs mass should be $\sim 314$ GeV, which is of course 
not consistent with the observed value. 
Let us assume that the SM is an effective field theory valid up to a scale $\Lambda$, with $\Lambda$ being the scale of new physics beyond the Standard Model. Then the Higgs mass radiative corrections of ${\cal O}(\Lambda^2)$ for the leading contributions, receives a huge fine-tuning mechanism to keep the renormalized mass as small as the measured Higgs mass or the electroweak scale \cite{Kolda:2000wi}.
Early attempts to get around the hierarchy problem was to expect new physics 
at the electroweak scale \cite{PhysRevD.14.1667,Veltman:1980mj,Witten:1981nf,WITTEN1981513,SHER1989273,Einhorn:1992um}. However, no new physics signal is detected so far at the electroweak scale, presumably demanding a reinterpretation of our notion for naturalness.
Before the discovery of the SM Higgs, the possibility that the Veltman condition can be satisfied at high scale while violated at low energy is studied in \cite{Jack:1989tv,Al-sarhi:1991gdi,Chaichian:1995ef}. There are also investigations indicating that the Veltman condition 
can be fulfilled close to the Planck scale \cite{Holthausen:2011aa,Hamada:2012bp,Jegerlehner:2013cta,Jegerlehner:2013nna}.

A prominent and ubiquitous mechanism to protect the Higgs mass correction from large amounts is through underlying symmetries, e.g. supersymmetric scenarios. This type of mechanism works if its dynamical scale is not large.
Another well known solution to the fine-tuning problem is the scale invariant models \cite{Plascencia:2015xwa,Kobakhidze:2014afa,Antipin:2013exa}.
The hierarchy problem is addressed in \cite{Masina:2013wja} in which the SM and its minimal extensions are merged with a high-scale supersymmetric theory at a scale where the Veltman condition is satisfied. 
The fine-tuning problem is discussed in the two-Higgs doublet model and left-right symmetric model in \cite{Boer:2019ipg,Biswas:2014uba,Chakraborty:2014oma,Jora:2013opa}. The Veltman condition is adapted to extended Higgs sectors in \cite{ElHedri:2018hix,Katz:2005au} to lower the fine-tuning of the SM Higgs mass.
In the same context there are studies with extended Higgs sector and various scenarios \cite{Kim:2018ecv,Chakraborty:2012rb,aali:2020tgr,Bian:2013xra,Karahan:2014ola,Chakrabarty:2018yoy,Darvishi:2017bhf,Chakraborty:2014xqa,Bazzocchi:2012pp,Ma:2014cfa,Abu-Ajamieh:2021vnh}. 

There are strong evidences that the SM cannot be an ultimate theory of particle physics and 
new physics might be at work at scales smaller than the Planck mass.
The size of the Higgs mass corrections may be controlled by adding new degrees of freedom in order to cause the cancellation of the radiative correction at the quadratic level at the scale $\Lambda \ll \Lambda_{\text{Pl}}$. An intriguing question is whether another weak scale paradigm, the WIMP (Weakly Interacting Massive Particle) scenario, would be interconnected to the naturalness problem intrinsically.
In the present work we exploit a dark matter model with two extra real singlet scalars both coupled to the SM Higgs. We consider a setup in which one of the singlet scalars is the 
dark matter (DM) candidate and the other singlet scalar is unstable. 
By running the couplings at high scale we study the hierarchy problem
and the effect of extra scalars on the Veltman conditions while respecting 
the stability of the extended Higgs potential.
There can be found various scenarios with the inclusion of two real singlet scalars emphasising on the dark matter or LHC phenomenology \cite{Abada:2011qb,Ahriche:2013vqa,Ghorbani:2014gka,DiazSaez:2021pfw,Claude:2021sye,Basak:2021tnj,McDowall:2019knq,Robens:2019kga,Ghorbani:2019itr,Bhattacharya:2017fid,Maity:2019hre}. 
As experimental bounds, we consider strong constraints from observed relic density granted by Planck and WMAP observations, bounds from direct detection experiments XENON1t and LUX and the invisible Higgs decay width. On the other hand, the theoretical constraints entail bounds from the perturbativity and the bounded from below condition (BFB condition) which we also call it the ``positivity condition'' in this article.

The setup of the paper is the following. 
In the next section we describe the two-scalar  model and impose bounds one the couplings from bounded from below (BFB) condition. Section \ref{phenomenology} is devoted to the phenomenology of the dark matter including bounds from the observed relic density, direct detection experiments and the invisible Higgs decay. The relevant Veltman conditions for the SM Higgs and the two singlet scalars are obtained in section \ref{Veltman}. The renormalization group equations and the numerical results for running of the couplings are given in section \ref{RGE}. We conclude our results in section \ref{conclusion}. In the Appendix we provide bounded from below conditions for a generic potential with two singlet scalars and N scalars added to the SM Higgs potential.

\section{Two-scalar model} 
\label{model}
The two real singlet scalar extension of the SM has the potential,
\begin{equation}\label{pot}
\begin{split}
 {\cal V_{\text{DM}}}  =& 
             \frac{1}{2} m^{2}_{1} \phi_{1}^2  +  \frac{1}{2} m^{2}_{2} \phi_{2}^2 
       +( \lambda_{H1} \phi_{1}^2 +  \lambda_{H2} \phi_{2}^2) H^{\dagger}H 
       + \lambda_H (H^{\dagger}H)^2 \\
       &+ \frac{1}{4} \lambda_1 \phi_{1}^4 + \frac{1}{4} \lambda_2 \phi_{2}^4 
       + \frac{1}{2} \lambda_{12} \phi_1^2 \phi_2^2 \,,
\end{split}
\end{equation} 
keeping only terms with dimensionless couplings. This potential is symmetric under 
$\mathbb{Z}_2$ 
such that under which $\phi_2 \to -\phi_2$ and $\phi_1 \to \phi_1$. 
According to this symmetry, terms like $\phi_1 \phi_2^3$ and $\phi_1^3 \phi_2$ 
are absent in the Lagrangian. 
The new singlet scalars interact with the SM Higgs beside having self-interactions. 
For the SM Higgs doublet we expand it around the minimum in the unitary gauge, 
$H^\dagger = (0‍ ~~ v+h')/\sqrt{2}$, and for the singlet scalars we assume
that only the scalar $\phi_1$ gets nonzero vacuum expectation value (VEV). Therefore, $\phi_1 = s^\prime_1 + w$ 
and $\phi_2 = s_2$ with  $\braket{\phi_1}=w$  and $\langle \phi_2 \rangle = 0$.
The dark matter candidate is then the scalar $s_2$. Moreover, the coupling $\lambda_{H2}$
is taken to be negligible at tree level, so we set $\lambda_{H2} = 0$. 
The running of the coupling $\lambda_{H2}$ will be studied by solving the renormalization group 
equations (RGE) in the next sections. We will also consider the case when $\lambda_{H2} \ne 0$. 
In order to find the physical mass eigenstates, $h$ and $s$, the following rotation is necessary in the space of the singlet scalars, $h'$ and 
$s^\prime_1$,
\begin{equation}
h = h' \cos \theta - s^\prime_1 \sin \theta \,, ~~ s_1 =  s^\prime_1 \cos \theta + h' \sin \theta \,, 
\end{equation}
where $\theta$ is the mixing angle. The physical masses corresponding to the scalar fields, $h, s_1$ and $s_2$ are  respectively $m_h, m_{s_1}$ and $m_{s_2}$. In the following we adapt $m_h = 125$ GeV and $m_{s_2} = m_{\text{DM}}$ being the mass of the Higgs and the DM mass respectively.
The quartic couplings $\lambda_H, \lambda_1$ and $\lambda_{H1}$ are obtained in terms of the mixing angle $\theta$ and the physical masses $m_h$ and $m_{s_1}$,

\begin{equation}
\label{coup}
\begin{split}
 \lambda_H =& \frac{1}{2v^2} (m_h^2 \cos^2 \theta + m_{s_1}^2 \sin^2  \theta) \\
 \lambda_1 =& \frac{1}{2w^2} (m_h^2 \sin^2 \theta + m_{s_1}^2 \cos^2  \theta) \\
 \lambda_{H1} =& \frac{\sin 2\theta}{4vw} (m_{s_1}^2- m_h^2) \,. 
\end{split}
 \end{equation}

The potential of the model must satisfy BFB condition (or the positivity condition) as a requirement for the vacuum stability. For the potential in Eq.~\ref{pot} the BFB condition becomes 
$\lambda_1>0 \wedge \lambda_2>0 \wedge \lambda_\text{H}>0$, and 

if  $ -\sqrt{\lambda_1 \lambda_2}<\lambda_{12}<\sqrt{\lambda_1 \lambda_2}$, then
\begin{equation}
\begin{split}
 & 
 \Big(-\sqrt{\lambda_\text{H}\lambda_1}  <\lambda_\text{H1}<-\lambda_{12}\sqrt{\lambda_\text{H}/\lambda_2} \\
 &~~~~~~~~~~~~~~~~~~~~~\wedge 
 \lambda_\text{H2}> \frac{\lambda_{12}\lambda_\text{H1}}{\lambda_1}-\frac{\sqrt{\lambda_1^2 \lambda_2 \lambda_\text{H}+\lambda_{12}^2 \lambda_\text{H1}^2-\lambda_1 \lambda_{12}^2\lambda_\text{H}-\lambda_1 \lambda_2 \lambda_\text{H1}^2}}{\lambda_1}  \Big)
  \\
& \vee  \left( 
 \lambda_\text{H1}> -\lambda_{12} \sqrt{\lambda_\text{H}/\lambda_2} \wedge \lambda_\text{H2}> -\sqrt{\lambda_2 \lambda_\text{H}} 
 \right) \,,
 \end{split}
\end{equation}

and if  $\lambda_{12}>\sqrt{\lambda_1 \lambda_2}$, then 

\begin{equation}
\lambda_\text{H1}>-\sqrt{\lambda_1 \lambda_\text{H}}  \wedge \lambda_\text{H2}> -\sqrt{\lambda_2 \lambda_\text{H}} \,.
\end{equation}
Setting $\lambda_\text{H2}=0$ in Eq.~\ref{pot} the BFB condition becomes $\lambda_1>0 \wedge \lambda_2>0 \wedge \lambda_\text{H}>0$, and 

if  $-\sqrt{\lambda_1 \lambda_2}<\lambda_{12}<0$, then
\begin{equation}
 \lambda_\text{H1}>-\sqrt{\frac{(\lambda_1 \lambda_2 - \lambda^2_{12})\lambda_\text{H}}{\lambda_2}} \,,
\end{equation}

and if  $\lambda_{12}>0$, then

\begin{equation}
 \lambda_\text{H1}>-\sqrt{\lambda_1\lambda_2} \,.
\end{equation}
When scanning the parameter space of the model the positivity conditions are imposed. 
The general conditions of the positivity for a potential with two scalars and N scalars are presented in the Appendix.

\section{Invisible Higgs decay, relic abundance and direct detection bounds}
\label{phenomenology}

We are interested in finding the viable regions in the coupling space for further use in the next sections. To this end, beside theoretical constraints, we impose additionally the constraints from WMAP \cite{Hinshaw:2012aka} and Planck \cite{Ade:2013zuv} observations on the DM relic 
density with $\Omega h^2 \sim 0.12$ and the constraints from the latest 
direct detection (DD) experimental bounds by XENON \cite{Aprile:2017iyp} and LUX \cite{Akerib:2016vxi}. 

\begin{figure}
\hspace{1.5cm}
\begin{minipage}{.2\textwidth}
\includegraphics{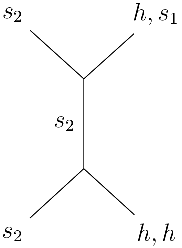}
\end{minipage}
\hspace{3cm}
\begin{minipage}{.2\textwidth}
\includegraphics{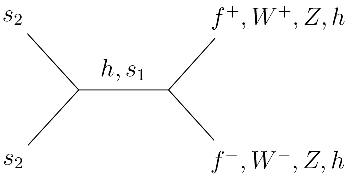}
\end{minipage} 
\caption{The relevant Feynman diagrams for dark matter annihilation at the freeze-out epoch.}
\label{anni-diagrams}
\end{figure}
We exploit the package {\tt MicrOMEGAs} \cite{Barducci:2016pcb} to compute the 
relic density and the DM-nucleus elastic scattering cross sections. 
The relevant Feynman diagrams for computation of the DM annihilation cross section is depicted in Fig.~\ref{anni-diagrams}. 
The DM particles annihilate into the SM particles ($h$,$W^\pm$,$Z$,$f$) 
through $s$-channel, and annihilate through $t$- and $u$-channel into two  SM Higgs ($hh$) 
and into $h s_1$.
Since it is assumed that the DM candidates are thermal relic, 
its number density, $n_2$, is governed by the Boltzmann equation,
\begin{equation}
 \frac{dn_{2}}{dt} +3Hn_{2} = - \langle \sigma_{\text{ann}}v_{\text{rel}} \rangle [n^{2}_{2}-(n^{\text{EQ}}_{2})^2 ]\,,
\end{equation}
where $H$ in the second term is the Hubble parameter. 
The thermal average of the annihilation cross section times the velocity at 
temperature $T$ is obtained by integration over the center of mass energy, 
\begin{equation}
\langle \sigma_{\text{ann}} v_{\text{rel}} \rangle = \frac{1}{8 m_{s_2}^4TK^{2}_{2}(\frac{m_{s_2}}{T})}
\int^{\infty}_{4m^{2}_{s_2}} ds~(s-4m^{2}_{s_2})\sqrt{s}~K_{1} (\frac{\sqrt{s}}{T})~\sigma_{\text{ann}}(s)\,,
\end{equation}
in which $K_1$ and $K_2$ are modified Bessel functions.
\begin{figure}
\begin{center}
\includegraphics{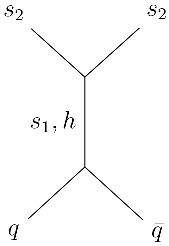}
\end{center}
\caption{The Feynman diagram for the DM direct detection scattering.}
\label{DD-diagrams}
\end{figure}
Moreover, DM candidates interact directly with normal matter through $t$-channel by exchanging a SM Higgs or the singlet scalar $s_1$, as shown in Fig.~\ref{DD-diagrams}. The DM elastic scattering cross section is spin-independent (SI) and is given by 
\begin{equation}
 \sigma_{\text{SI}} = \frac{1}{4\pi} \frac{m_N^4 f_N^2}{(m_N+m_{s_1})^2} \frac{w^2 \lambda_{12}^2\sin 2\theta}{v^2} (\frac{1}{m_h^2}-\frac{1}{m_{s_1}^2})^2 \,,
\end{equation}
where $m_N = 0.938$ GeV is the nucleon mass and the scalar form-factor is $f_N = 0.3$.
When DM mass is small such that $m_{s_2} < m_h/2$, then there are limits from invisible Higgs decay. The ATLAS experiment provides upper limits for the invisible branching ratio
as $\text{Br}^{\text{exp}}_{\text{inv}} < 0.26$ at $95\%$ C.L.\cite{ATLAS:2019cid}, and the CMS experiment gives a stronger limit as $\text{Br}^{\text{exp}}_{\text{inv}} < 0.19$ at $95\%$ C.L \cite{CMS:2018yfx}. In this work we take the stronger limit from the CMS. The decay width of the SM Higgs is 
$\Gamma_{\text{SM}} = 4.07$ MeV \cite{ParticleDataGroup:2020ssz}. When Higgs decay to the singlet scalar $s_1$ is kinematically allowed then the total decay width modifies as
$\Gamma_{\text{SM}} \cos^2 \theta +\Gamma_{\text{inv}}$. 
Given the result from the CMS, the theoretical invisible decay width should 
satisfy the relation $\Gamma_{\text{inv}} < 0.95 \cos^2 \theta$ MeV.
The invisible Higgs decay through the process $H \to s_1 s_1$ is given by
\begin{equation}
 \Gamma_{\text{inv}} = \frac{\lambda_{12} w^2 \sin^2 \theta}{8\pi m_h} \sqrt{1-4 m_{s_1}^2/m_h^2} \,.
\end{equation}
\begin{figure}
\begin{minipage}{.25\textwidth}
\includegraphics[width=1.5\textwidth,angle =-90]{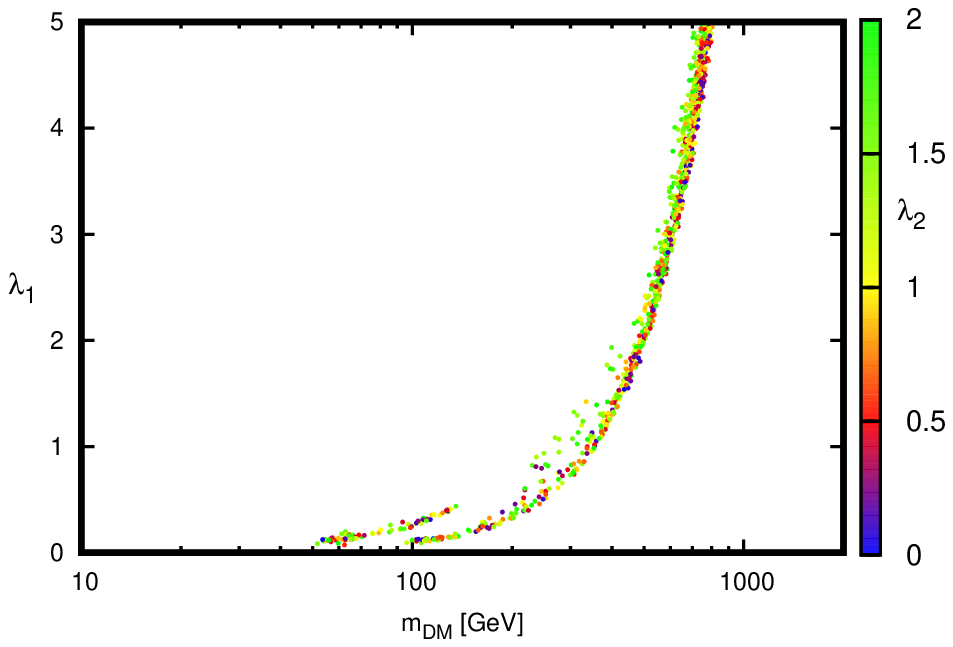}
\end{minipage}
\hspace{4cm}
\begin{minipage}{.25\textwidth}
\includegraphics[width=1.5\textwidth,angle =-90]{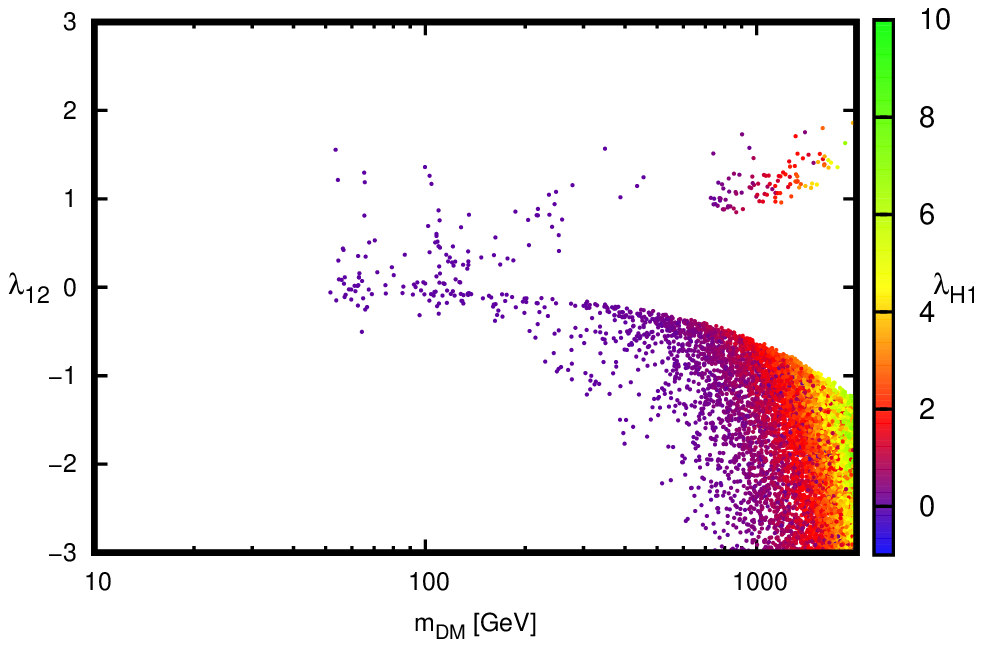}
\end{minipage} 
\caption{The viable region restricted by the WMAP/Planck observed relic density
and direct detection bounds from LUX/XENON1t for the scalar dark matter 
mass in the range $10~\text{GeV}$-$2000~\text{GeV}$. 
The BFB conditions are imposed on the couplings. Here we took $\lambda_{\text{H2}} = 0$}
\label{couplings-scan1}
\end{figure}
As free parameters of the model, we choose, $m_{\text{DM}}$, $\delta$, 
$\theta$, $\lambda_{12}$, $\lambda_{2}$, $\lambda_{\text{H2}}$ and $w$, 
where $\delta = m_{s_1}-m_{\text{DM}}$.  
For the computation in this section we fix $w = 250$ GeV.
First, we set $\lambda_{\text{H2}} =0$ and scan over the parameter space within these ranges, 
$0.969<\cos \theta<1$, $0 <\lambda_{2}<2$, $-\sqrt{\lambda_1 \lambda_2} < \lambda_{12} < 2$ ,$10~\text{GeV} < m_{\text{DM}}<2~$TeV and $0 <\delta <100$ GeV. 
The couplings $\lambda_{H}$, $\lambda_{1}$ and $\lambda_{\text{H1}}$ are given in 
terms of the free parameters in Eq.~\ref{coup}.
In Fig.~\ref{couplings-scan1} the viable scalar couplings are shown after placing constraints from the BFB conditions, the observed relic density and the DD upper bounds.   
Next, we relax the condition of taking $\lambda_{\text{H2}}$ negligible and in our scan we choose $-1< \lambda_{\text{H2}}< 1$. Respecting the ranges discussed earlier for the parameters,
we present the limits on the couplings in Fig.~\ref{couplings-scan2}
 after imposing the BFB condition, the observed relic density and the DD bounds. 

The results found here are beyond the expectation of the model being simply a duplication of the singlet scalar model. In the singlet scalar model, the DM annihilation amplitude and the elastic scattering amplitude in the DM-nucleon collision, share the same interaction coupling for the entire DM mass range. This property makes it hard to get small direct detection cross section and at the same time large annihilation cross section needed for the correct DM relic density. 
The situation is changed significantly in the two real singlet model. 
This point is discussed in \cite{Ghorbani:2014gka,Claude:2021sye} for various setups.
Beside having strong constraints on the couplings from the BFB conditions of the potential there exist a large parameter space respecting both the observed DM density and the DD bounds.

\begin{figure}
\label{couplings-scan2}
\begin{minipage}{0.25\textwidth}
\includegraphics[width=1.5\textwidth,angle =-90]{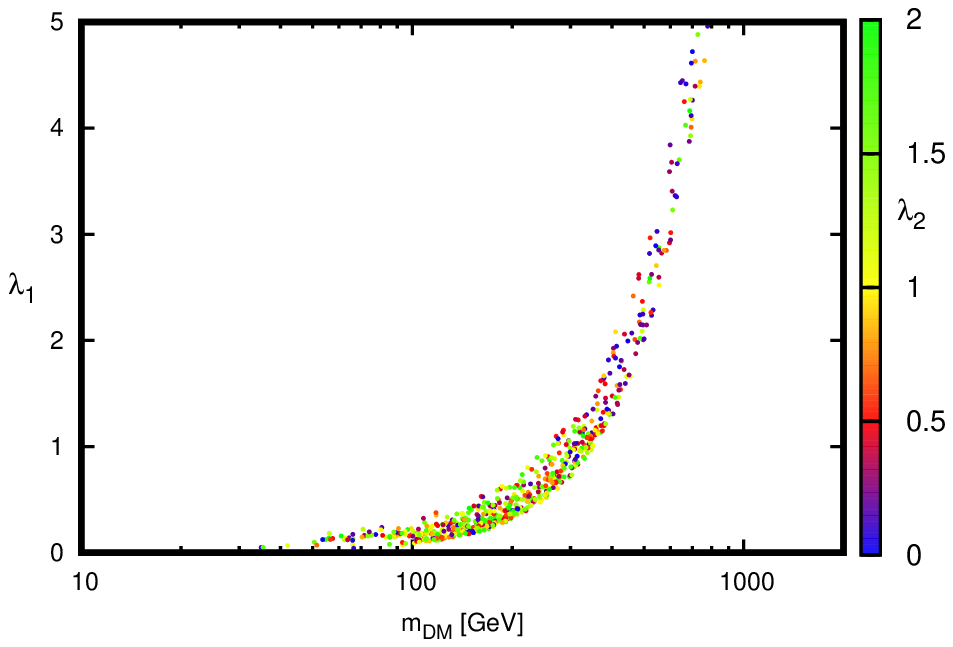}
\end{minipage}
\hspace{4cm}
\begin{minipage}{0.25\textwidth}
\includegraphics[width=1.5\textwidth,angle =-90]{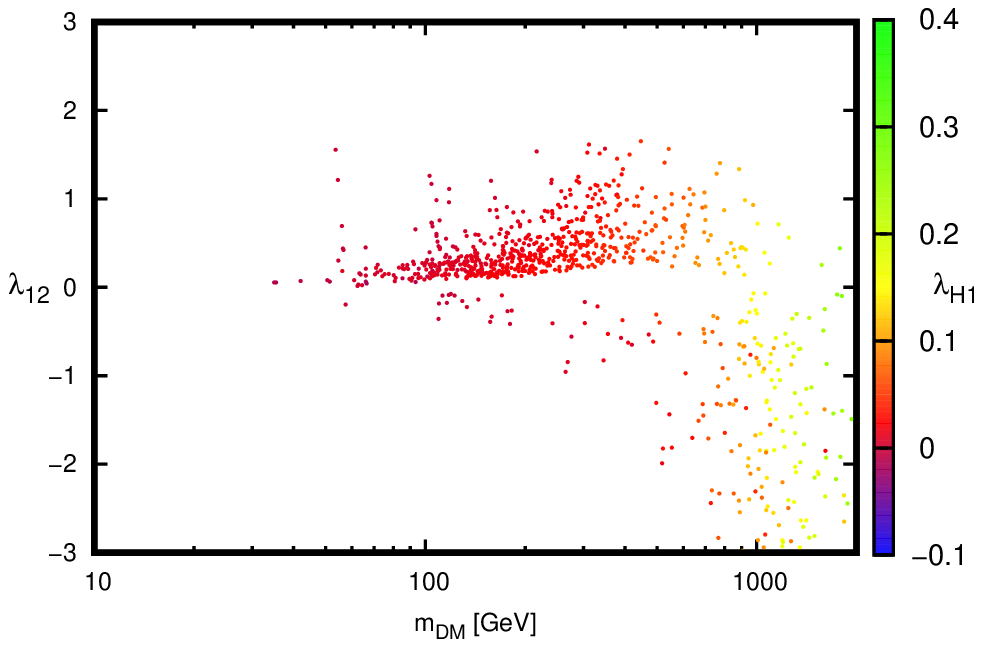}
\end{minipage}
\caption{Same as Fig.~\ref{couplings-scan1} but with $\lambda_{H2} \ne 0$.}
\label{couplings-scan2}
\end{figure}

\section{Veltman parameters} 
\label{Veltman}
To obtain the mass correction of the scalars we rewrite the Lagrangian 
in terms of the fields in the 
mass eigenstates. 
The Higgs mass correction at one loop keeping only quadratic 
correction in the mass scale $\Lambda$, incorporates the new 
scalar couplings $\lambda_{H1}$, 
$\lambda_{H2}$, $\lambda_1$ and $\lambda_{12}$ as,
\begin{equation}
\begin{split}
\delta m_h^2 =& \frac{\Lambda^2}{16\pi^2} \Big[ 6\lambda_H \cos^2 \theta  
 +6 \lambda_1 \sin^2 \theta  + 12 \lambda_{H1} \sin^2 \theta \cos^2 \theta   + 
    (\frac{3}{4} g_{1}^2 + \frac{9}{4} g_{2}^2 - 4 \lambda_t^2) \cos^2 \theta \\
    &+ 2 \lambda_{H1} + 2 \lambda_{H2} \cos^2 \theta - 12 \lambda_{H1} \sin^2 \theta \cos^2 \theta
    +6 \lambda_1 \sin^2 \theta \cos^2 \theta \\
    &+6 \lambda_H \sin^2 \theta \cos^2 \theta + 2 \lambda_{12}  \sin^2 \theta \Big] 
       \equiv  \frac{\Lambda^2}{16\pi^2} V_H \,.
\end{split}
\end{equation}
The measurement of the Higgs boson signal strengths \cite{ParticleDataGroup:2020ssz}
constrains the mixing angle to small values, $|\cos \theta| > 0.969$. 
Therefore, we will approximate the Veltman parameter $V_{H}$, at small mixing angles,
\begin{equation}
 V_{H} \sim  6 \lambda_H + 2 \lambda_{H1} + 2 \lambda_{H2} + \frac{3}{4} g_{1}^2 
 + \frac{9}{4} g_{2}^2 -  4 \lambda_t^2 \,.
\label{VeltmanParameter}
 \end{equation}
In the same way the singlet scalar masses get quadratic corrections as ,
 \begin{equation}
  \begin{split}
  \delta m^2_{s_1} & = \frac{\Lambda^2}{16\pi^2} \Big[ 6 \lambda_H \sin^4 \theta + 6 \lambda_1 \cos^4 \theta  
  + 12 \lambda_{H1} \sin^2 \theta \cos^2 \theta + 2 \lambda_{H2} \sin^2 \theta \\
  &+ 2 \lambda_{12} \cos^2 \theta + 6 \lambda_H \sin^2 \theta \cos^2 \theta  
    + 6 \lambda_1 \sin^2 \theta \cos^2 \theta \\
   &- 6 \lambda_{H1} \sin^2 \theta \cos^2 \theta +2 \lambda_{H1} \Big] 
   \equiv  \frac{\Lambda^2}{16\pi^2} V_{s_1} \,, \\
  \delta m^2_{s_2} & = \frac{\Lambda^2}{16\pi^2} \Big[ 2 \sin^2 \theta \lambda_{H2} + 2 \lambda_{12}   + 2 \cos^2 \theta \lambda_{H2} +  6 \lambda_{2} \Big] 
  \equiv  \frac{\Lambda^2}{16\pi^2} V_{s_2} \,. 
  \end{split}
 \end{equation}
At the limit of small mixing angle we arrive at 
 \begin{equation} 
 \label{VeltmanSinglet}
 \begin{split}
    V_{s_1} \sim    
                      2 \lambda_{12}  + 2 \lambda_{H1} + 6 \lambda_1   
     \,, \\
 V_{s_2}    \sim      2 \lambda_{12} + 2 \lambda_{H2} +  6 \lambda_2    
    \,.  
  \end{split}  
  \end{equation}  
A comment is appropriate to mention here. 
We have used the cutoff regularization in computing radiative 
corrections of the Higgs mass and the singlet scalar masses.
The cutoff regularization has an interesting property of 
disentangling quadratic and logarithmic divergences. On the other hand, in the 
dimensional regularization there is no discrimination between these two types of 
divergences and one may get a somewhat different quantum corrections as discussed in \cite{Einhorn:1992um}.
Since in the Veltman approach the goal is to cancel the strongest divergences (the quadratic divergences) it is sufficient to apply the cutoff regularization \cite{Chakraborty:2012rb,Masina:2013wja}.

The corresponding Veltman conditions at some high energy scale, $\Lambda$, 
are $V_H(\Lambda) \sim 0$, $V_{s_1}(\Lambda) \sim 0$ 
and $V_{s_2}(\Lambda) \sim 0$.
Now by imposing $V_H(\Lambda) = 0 $, from Eq.~\ref{VeltmanParameter} we obtain 
\begin{equation}
   \lambda_{H2} + \lambda_{H1} = 2 \lambda_t^2 - 3 \lambda_H  - \frac{3}{8} g_{1}^2 
 - \frac{9}{8} g_{2}^2 \,. 
\end{equation}
At the same scale we impose $V_{s_1}(\Lambda) = 0$ and $V_{s_2}(\Lambda) = 0$, and then 
from Eq.~\ref{VeltmanSinglet}, we arrive at 
\begin{equation}
 \begin{split}
  \lambda_{12} &= - \lambda_{H1} -3 \lambda_{1} \,, \\
 \lambda_{2} &= \lambda_{1} + \frac{1}{3} \lambda_{H1} - \frac{1}{3} \lambda_{H2} \,.
\end{split}
\end{equation}
We recall the relations in Eq.~\ref{coup} for the couplings, $\lambda_{H}$, $\lambda_{H1}$ and $\lambda_{1}$.

\section{Renormalization group equations and Veltman conditions}
\label{RGE}
The running of the couplings are controlled by the $\beta$ functions in the renormalization group equations. The $\beta$ function for the coupling $\lambda$ is defined as
\begin{equation}
 \beta_{\lambda} = \frac{d\lambda}{d\ln(\mu/\mu_0)} \,,
\end{equation}
where $\mu$ is the renormalization scale with the initial value $\mu_0$.
In this work the $\beta$ functions are computed at one loop order. 
The gauge couplings, $g_1, g_2$ and $g_3$, remain intact and are similar to those in the SM, 
\begin{equation}
 16\pi^2 \beta_{g_i} = b_i g_{i}^3 \,,
\end{equation}
where $b_{i} = \frac{41}{6}, -\frac{19}{6}, -7$. There are exact formulas for the 
running of the gauge couplings,
\begin{equation}
 g_i(\mu) = \frac{g_i(\mu_0)}{ \Large(1-2b_i g_i^2 \ln(\mu/\mu_0)\Large)^{1/2}} \,.
\end{equation}
The $\beta$ function for the top quark Yukawa coupling at one loop is also the same as the SM one,
\begin{equation}
 16\pi^2 \beta_{\lambda_t} = \lambda_{t} (\frac{9}{2} \lambda_t^2 -\frac{17}{12} g_1^2 
                                   - \frac{9}{4} g_2^2 -8 g_3^2) \,.
\end{equation}
The $\beta$ functions for the new dimensionless couplings in the potential and also for the 
quartic Higgs self-coupling at one loop are as follows\footnote{We have checked our results applying the package SARAH \cite{Staub:2012pb}.},
\begin{equation}
\begin{split}
 16\pi^2 \beta_{\lambda_H} = &  \frac{3}{8} g_1^4 + \frac{3}{4} g_1^2 g_2^2 + \frac{9}{8} g_2^4
  - 3 g_1^2 \lambda_H  - 9 g_2^2 \lambda_H \\
  &+ 24 \lambda_H^2 + 2 \lambda_{H1}^2 + 2 \lambda_{H2}^2 
    +12 \lambda_H \lambda_t^2 -6 \lambda_t^4 \,, \\
16\pi^2 \beta_{\lambda_{H1}} =& -\frac{3}{2} g_1^2 \lambda_{H1} -\frac{9}{2} g_2^2 \lambda_{H1}
                             + 2 \lambda_{12} \lambda_{H2} +6 \lambda_1 \lambda_{H1} \\
                             &+ 12 \lambda_H \lambda_{H1} 
                             + 8 \lambda_{H1}^2  + 6 \lambda_t^2 \lambda_{H1} \,, \\
16\pi^2 \beta_{\lambda_{H2}} =& -\frac{3}{2} g_1^2 \lambda_{H2} -\frac{9}{2} g_2^2 \lambda_{H2}
                             + 2 \lambda_{12} \lambda_{H1} + 6 \lambda_2 \lambda_{H2} \\
                             &+ 12 \lambda_H \lambda_{H2}                             + 8 \lambda_{H2}^2 + 6 \lambda_t^2 \lambda_{H2} \,, \\
16\pi^2 \beta_{\lambda_{1}} =&  18 \lambda_1^2 + 2 \lambda_{12}^2 + 8 \lambda_{H1}^2 \,, \\
16\pi^2 \beta_{\lambda_{2}} =& 18 \lambda_2^2 +2 \lambda_{12}^2 + 8 \lambda_{H2}^2 \,,\\
16\pi^2 \beta_{\lambda_{12}} = & 8 \lambda_{12}^2 + 6 \lambda_1 \lambda_{12} + 6 \lambda_2 \lambda_{12}+ 8 \lambda_{H1} \lambda_{H2} \,,
 \end{split}
\end{equation}
\begin{figure}
\begin{center}
\includegraphics[width=.65\textwidth]{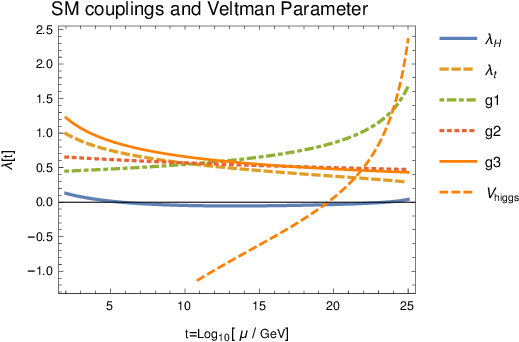}
\caption{The running of all the SM couplings and Higgs Veltman parameter are shown.}
\end{center}
\label{running-SMcouplings}
\end{figure}
where we have set all the SM Yukawa couplings equal to zero and only retained the 
top quark Yukawa coupling, since the light quark and lepton Yukawa couplings 
are expected to have quite small effects in the runnings.
We first show the running of the SM couplings in the 
set $\{g_1,g_2,g_3,\lambda_t,\lambda_{H}\}$ in Fig.~\ref{running-SMcouplings}.
As it is evident, the coupling $\lambda_{H}$ tends to zero at $\Lambda \sim 10^5$ GeV and therefore violates the bounded from below condition of the potential. 
In addition we take a look at the running of the Veltman parameter, $V_H$, in Fig.~\ref{running-SMcouplings} and notice that only at 
$\Lambda \sim 10^{20}$ GeV, the Veltman condition is fulfilled. 
This scale is larger than the Planck scale.  
Next, we consider two representative benchmarks in the viable parameter space in the two-scalar model. We have found a point at the electroweak scale (at the top quark mass) which respects all the constraints discussed earlier in the paper, with the following parameters,\\

$\bullet$ Benchmark I: \\

$m_{\text{DM}} = 431 $ GeV, $m_{s_1} = 505 $ GeV, $\lambda_H = 0.42$, 
$\lambda_{H1}= -0.021$, $\lambda_{H2}= -0.02$, 
$\lambda_1 = 9.9 \times 10^{-3} $, $\lambda_2= 7.1 \times 10^{-3}$, $\lambda_{12}= -6\times 10^{-4}$  , $\sin \theta = -0.21$, $w =3587$ GeV. \\

\begin{figure}
\hspace{-1cm}
\begin{minipage}{0.28\textwidth}
\includegraphics[width=2\textwidth]{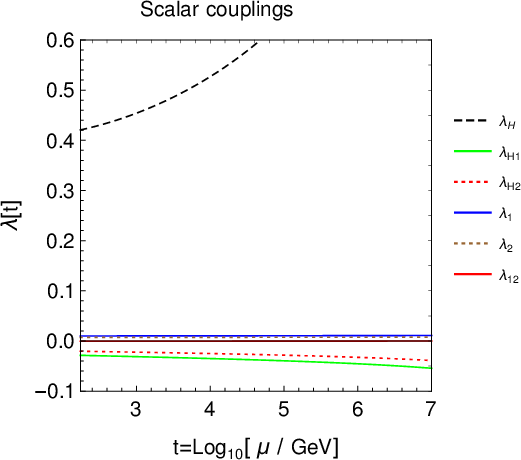}
\end{minipage}
\hspace{4.1cm}
\begin{minipage}{0.27\textwidth}
\includegraphics[width=1.7\textwidth]{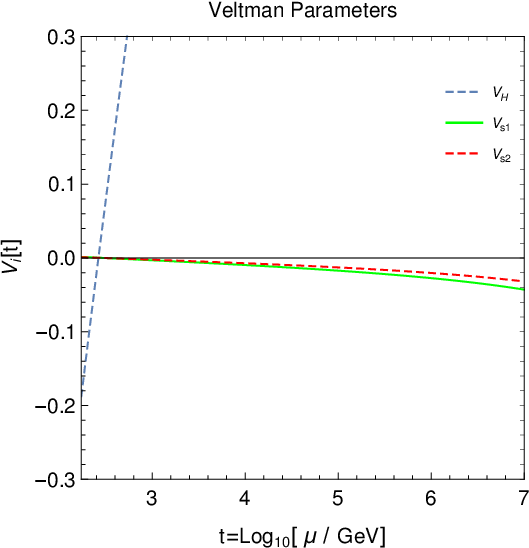}
\end{minipage}
\caption{In the {\it left panel} the running of all the scalar couplings, and in the {\it right panel} the running of the Veltman parameters are shown for the Benchmark-I in the parameter space which fulfills the constraints from observed relic abundance, direct detection and theoretical bounds. The Veltman parameters are vanishing at $300$ GeV.}
\label{bench1}
\end{figure}

As can be seen from the plots in Fig.~\ref{bench1}, the Veltman conditions are satisfied at the scale $\sim 300$ GeV. Following the argumentation in \cite{Barbieri:1987fn,Kolda:2000wi},
the measure of fine-tuning as a function of $\Lambda$ is given by 
\begin{equation}
 {\cal F} \equiv \frac{|\delta m_h^2|}{m_h^2} = \frac{\Lambda^2}{16 \pi^2 m_h^2} |V_H(\Lambda)| \,.
\end{equation}
Then, we would say that the electroweak scale is fine-tuned to one part in 
${\cal F}$, and ${\cal F} \leq 1$ indicates the absence of tuning.
For the representative point at the electroweak scale $\Lambda \sim 173$ GeV provided above, we obtain ${\cal F} \sim 0.002$, which means no tuning at the electroweak scale. 

Our second Benchmark point with the following parameters, satisfies the Veltman conditions at higher scale,\\

$\bullet$ Benchmark II: \\

$m_{\text{DM}} = 424 $ GeV, $m_{s_1} = 828 $ GeV, $\lambda_H = 0.36$, 
$\lambda_{H1}= -0.039$, $\lambda_{H2}= -0.025$, 
$\lambda_1 = 9.8 \times 10^{-3} $, $\lambda_2= 5.5 \times 10^{-3}$, $\lambda_{12}= 
0.015$  , $\sin \theta = -0.2$, $w =5916$ GeV. \\

\begin{figure}
\hspace{-1cm}
\begin{minipage}{0.28\textwidth}
\includegraphics[width=2\textwidth]{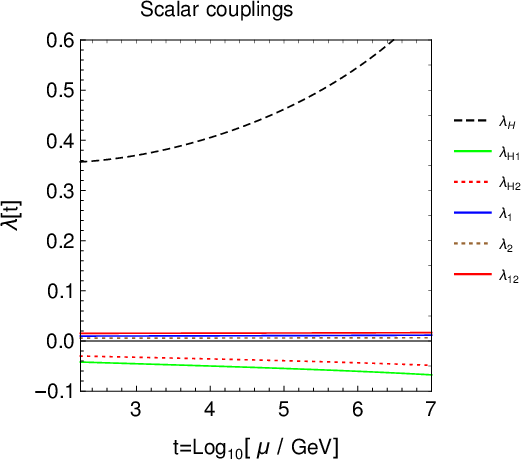}
\end{minipage}
\hspace{4.1cm}
\begin{minipage}{0.27\textwidth}
\includegraphics[width=1.7\textwidth]{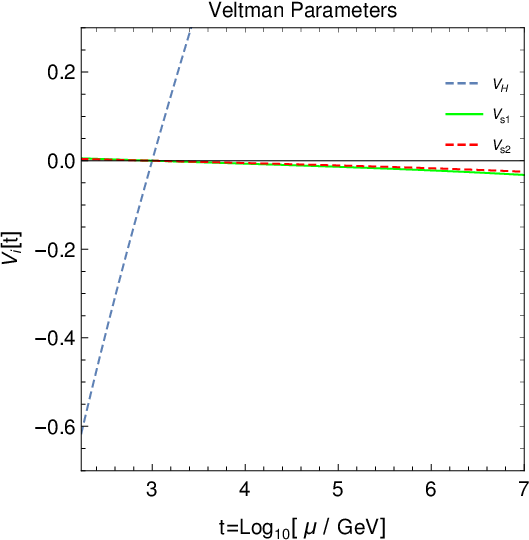}
\end{minipage}
\caption{In the {\it left panel} the running of all the scalar couplings, and in the {\it right panel} the running of the Veltman parameters are shown for the Benchmark-II in the parameter space which fulfills the constraints from observed relic abundance, direct detection and theoretical bounds. The Veltman parameters are vanishing at $1$ TeV.}
\label{bench2}
\end{figure}

The running of the couplings and the Veltman parameters are shown in Fig.~\ref{bench2}.
The Veltman conditions are satisfied for all scalars at 1 TeV. For this Benchmark point we find 
${\cal F} \sim 0.007$ at $\Lambda \sim 173$ GeV, which again 
indicates no tuning at the electroweak scale. 

\section{Conclusion}
\label{conclusion}
In this work we considered a dark matter model with two real singlet scalars. The lighter scalar is taken as the DM candidate. 
We have found regions in the parameter space which explains correctly the observed DM relic density. In contrast to the singlet scalar DM model, the two-scalar DM model
evades the latest bounds from LUX and XENON1t direct detection experiments. Furthermore,  we aimed at considering also the vacuum stability and the hierarchy problem within the Veltman approach.  
Upon running of the couplings by solving the renormalization group equations, we consider the points in the parameter space which respect both the vacuum stability and the perturbativity conditions. Another theoretical restriction is to find points which render the Veltman conditions satisfied at energies much lower than the Planck scale. As benchmark-I and benchmark-II we have shown that the Veltman parameters for all scalars in the model including the Higgs are vanishing at $300$ GeV and $1$ TeV respectively. The fine-tuning parameter for these benchmarks is shown to be much smaller than one, resulting in models with no fine-tuning problem. 
We conclude that adding more singlet scalars can resolve different fundamental problems in the SM including the perturbativity and the vacuum stability problem in the SM (as shown in \cite{Ghorbani:2021rgs}), the DM relic density while evading all related bounds, and the fine-tuning problem via the Veltman approach.

\appendix

\section{Bounded from below condition (positivity)}\label{Apen}

 In this Appendix we follow \cite{Kannike:2016fmd}. Let us write the generic two-scalar quartic potential $V_4$ as, 
 
 \begin{equation}\label{V4}
  \mathcal{V}_4= \lambda_\text{H} |H|^4 +  M(\phi_1,\phi_2) |H|^2 + V(\phi_1,\phi_2),
 \end{equation}
where, 
\begin{equation}\label{M}
 M(\phi_1,\phi_2)=  \lambda_\text{H1} \phi_1^2+  \lambda_\text{H2} \phi_2^2 + 2 \lambda_\text{H12} \phi_1 \phi_2,
\end{equation}
and 
\begin{equation}\label{V}
 V(\phi_1,\phi_2)= \frac{1}{4}\lambda_1 \phi_1^4 +\frac{1}{4} \lambda_2 \phi_2^4 + \frac{1}{2}\lambda_{12} \phi_1^2 \phi_2^2 + \lambda_{13} \phi_1 \phi_2^3 +\lambda_{31} \phi_1^3 \phi_2
\end{equation}
The BFB condition (positivity condition) $\mathcal{V}_4>0$ immediately holds if, 
\begin{equation}\label{con1}
 \lambda_\text{H}>0 ~~ \wedge ~~ M(\phi_1,\phi_2)>0 ~~ \wedge ~ V(\phi_1,\phi_2)>0.
\end{equation}
Alternatively we can write $\mathcal{V}_4=(2\lambda_\text{H}|H|^2+M^2)^2/4\lambda_\text{H}+V_\text{min}$ where $\mathcal{V}_\text{min}=V-M^2/4\lambda_\text{H}$ is the value of $\mathcal{V}_4$ at the minimum. Therefore the positivity condition becomes, 
\begin{equation}\label{con2}
 \lambda_\text{H}>0 ~~ \wedge  ~~ \mathcal{V}_\text{min}\equiv V(\phi_1,\phi_2)-\frac{M^2(\phi_1,\phi_2)}{4\lambda_\text{H}}>0.
\end{equation}
Therefore the positivity condition $\mathcal{V}_4>0$ is given either by Eq. (\ref{con1}) or Eq. (\ref{con2}). 

We still need to restrict the parameters with conditions $M>0$, $V>0$ in Eq. (\ref{con1}) and $V_\text{min}>0$ in Eq. (\ref{con2}). Making a rotation in $(\phi_1,\phi_2)$ space, the polynomial $M$ can be diagonalized: $M(\phi_1',\phi_2')= \lambda_\text{H1}' \phi_1'^2+  \lambda_\text{H2}' \phi_2'^2$ where $\lambda_\text{H1}'$ and $\lambda_\text{H2}'$ are eigenvalues of the Hessian matrix for $ M(\phi_1,\phi_2)$. Therefore, $M(\phi_1,\phi_2)>0$ is equivalent to, 
\begin{subequations}
 \begin{align}
  \lambda_\text{H1}'\equiv  \lambda_\text{H1} + \lambda_\text{H12} -\sqrt{(\lambda_\text{H1} -\lambda_\text{H12})^2+4 \lambda_\text{H2} \lambda_\text{H12}}>0\,,\\
  \lambda_\text{H1}'\equiv  \lambda_\text{H1} + \lambda_\text{H12} +\sqrt{(\lambda_\text{H1} -\lambda_\text{H12})^2+4 \lambda_\text{H2} \lambda_\text{H12}}>0\,.
 \end{align}
\end{subequations}
The condition $V>0$ turns into the question of positivity of a quartic polynomial if we divide $V$ by $\phi_1^4$ or $\phi_2^4$, e.g., $V/\phi_2^4\equiv f(x)=a x^4 +b x^3 + c x^2 + d x + e
> 0$ with $x\equiv \phi_1^4/\phi_2^4$, and 
\begin{equation}\label{Vabcde}
 a=\frac{\lambda_1}{4}~~~ b=\lambda_{31}~~~ c=\frac{\lambda_{12}}{2}~~~ d=\lambda_{13}~~~ e=\frac{\lambda_2}{4}\,.
\end{equation}
At $x=0$ and $x\to \infty$ limits, the positivity condition requires $\lambda_1>0$ and $\lambda_2>0$. Then $f(x)$ remains positive if it never intersects the real axis. In other words, the positivity holds when the roots are always complex. The complexity of the roots is guaranteed by the conditions,
\begin{equation}\label{noroot}
 \Delta>0 \wedge (P>0 \vee D>0)
\end{equation}
where $\Delta$ is the discriminant, 
\begin{equation}\label{Delta}
 \begin{split}
  \Delta &=256 a^3 e^3-192 a^2 b d e^2-128 a^2 c^2 e^2+144 a^2 c d^2 e-27 a^2 d^4+144 a b^2 c e^2\\
  &-6 a b^2 d^2 e-80 a b c^2 d e+18 a b
   c d^3+16 a c^4 e-4 a c^3 d^2-27 b^4 e^2+18 b^3 c d e\\
   &-4 b^3 d^3-4 b^2 c^3 e+b^2 c^2 d^2\,,
 \end{split}
\end{equation}
and 
\begin{equation}\label{PD}
 P= 8 a c - 3b^2, ~~~D = 64 a^3 e - 16 a^2 c^2 + 16 a b^2 c - 16 a^2 b d - 3 b^4\,.
\end{equation}
The same approach can be used to find regions with $V_\text{min}>0$. From (\ref{con2}) $V_\text{min}=a x^4+b x^3+c x^2 + d x +e$ with
\begin{equation}\label{Vminabcde}
\begin{split}
 a= &\frac{\lambda_1}{4}-\frac{\lambda_\text{H1}^2}{4\lambda_\text{H}},~~~ b= \lambda_{31}-\frac{\lambda_\text{H1}\lambda_\text{H12}}{\lambda_\text{H}},~~~ c= \frac{\lambda_{12}}{2}-\frac{2\lambda_\text{H12}^2+\lambda_\text{H1}\lambda_\text{H2}}{2\lambda_\text{H}},\\
 &d= \lambda_{13}-\frac{\lambda_\text{H2}\lambda_\text{H12}}{\lambda_\text{H}},~~~ e= \frac{\lambda_2}{4}-\frac{\lambda_\text{H2}^2}{4\lambda_\text{H}}\,.
 \end{split}
\end{equation}
These parameters again must satisfy the conditions in Eq. (\ref{noroot}) with $\Delta$, $P$ and $D$ evaluated using Eq. (\ref{Vminabcde}).

\subsection{Biquadratic potential}
For biquadratic potential or in the absence of the odd terms in Eq. (\ref{pot}), i.e. by setting $\lambda_{H12}=\lambda_{13}=\lambda_{31}=0$, the positivity conditions will be determined as before, with $b$ and $d$ being vanishing in the parameters $\Delta$, $P$, and $D$ given in Eqs. (\ref{Delta}) and (\ref{PD}). Therefore, in biquadratic potentials we have,
\begin{equation}
 \Delta =256 a^3 e^3-128 a^2 c^2 e^2 +16 a c^4 e,~~~P= 8 a c, ~~~D = 64 a^3 e - 16 a^2 c^2
\end{equation}
The condition $V>0$ is given by Eq. (\ref{noroot}) with,
\begin{equation}
 a=\frac{\lambda_1}{4},~~~ c=\frac{\lambda_{12}}{2},~~~ e=\frac{\lambda_2}{4}\,,
\end{equation}
  and $V_\text{min}>0$ with,
  \begin{equation}\label{Vminabcde2}
 a= \frac{\lambda_1}{4}-\frac{\lambda_\text{H1}^2}{4\lambda_\text{H}},~~~  c= \frac{\lambda_{12}}{2}-\frac{\lambda_\text{H1}\lambda_\text{H2}}{2\lambda_\text{H}},~~~ e= \frac{\lambda_2}{4}-\frac{\lambda_\text{H2}^2}{4\lambda_\text{H}}\,.
\end{equation} 

\subsection{Positivity in biquadratic potential with N scalars}
Let us consider a generic potential as  $\mathcal{V}_4= 1/2 \lambda_{ij} \phi_i^2  \phi_j^2$ where $i,j=1,...,N$ with the convention that $\lambda_{ii}\equiv \lambda_i /2$. When only $\phi_1\to \infty$ while other fields are finite, we need to impose the condition $\lambda_{11}\equiv \lambda_1>0$. When two scalars are large, say $\phi_1\to \infty, \phi_2\to \infty$ and the rest of the scalars are finite, the positivity imposes,
\begin{equation}
 \lambda_1>0  ~~~~~\wedge~~~~~ \lambda_1 \lambda_2 -\lambda_{12}^2>0
\end{equation}
This trend should continue until N scalars. The Hessian matrix $\mathcal H_{ij}\equiv \partial^2 V_4/\partial \phi^2_i \partial \phi^2_j$ in general can be written as
\begin{equation}
\mathcal H=
\begin{pmatrix}
\lambda_{11} & \lambda_{12} &\dots& \lambda_{1N}\\
\lambda_{21} & \lambda_{22} &\dots& \lambda_{2N} \\
\vdots & \vdots & \ddots & \vdots \\
\lambda_{N1} & \lambda_{N2} &\dots& \lambda_{NN}\\
\end{pmatrix}
\end{equation}
that is a symmetric matrix, i.e. $\lambda_{ij}=\lambda_{ji}$. The positivity condition of $N\times N$ Hermitian matrix (symmetric matrix in our case) {\it \`{a} la } Sylvester is the positivity of determinants for all principal minors of the matrix $\mathcal H$, 
\begin{equation}
 \lambda_1>0 ~~~~~\wedge~~~~~
  \begin{vmatrix}
\lambda_{1} & \lambda_{12}\\
\lambda_{21} & \lambda_{2}\\
\end{vmatrix}>0  ~~~~\wedge~~~~ \dots ~~~~\wedge~~~~
\begin{vmatrix}
\lambda_{1} & \lambda_{12} &\dots& \lambda_{1N}\\
\lambda_{12} & \lambda_{2} &\dots& \lambda_{2N} \\
\vdots & \vdots & \ddots & \vdots \\
\lambda_{1N} & \lambda_{2N} &\dots& \lambda_{N}\\
\end{vmatrix}>0 \,.
\end{equation}

\bibliography{ref.bib}
\bibliographystyle{utphys}
\end{document}